\documentclass[letterpaper,conference]{IEEEtran}
\usepackage{cite}
\usepackage{graphicx}
\usepackage{amsmath}
\usepackage{enumerate}
\usepackage{amsthm}
\usepackage[linesnumbered,lined,boxed,commentsnumbered,ruled,vlined]{algorithm2e}
\usepackage[noend]{algpseudocode}
\DeclareMathOperator*{\argmax}{arg\,max}

\makeatletter
\def\BState{\State\hskip-\ALG@thistlm}
\makeatother
\usepackage{color}
\theoremstyle{definition}

\usepackage{amssymb}
\IEEEoverridecommandlockouts
\begin{document}
	
	\title{Network-Assisted D2D Relay Selection Under the Presence of Dynamic Obstacles *\thanks{\noindent * A preliminary version of this paper has been accepted for publication in the proceedings of the 44th IEEE Conference on Local Computer Networks (LCN), October 14-17, 2019, Osnabr\"{u}ck, Germany.}}
	\author{\IEEEauthorblockN{Durgesh Singh and Sasthi C. Ghosh\\}
		\IEEEauthorblockA{Advanced Computing \&  Microelectronics Unit,\\ Indian Statistical Institute, Kolkata 700108, India\\
			Email: durgesh.ccet@gmail.com, sasthi@isical.ac.in}}
	\maketitle

	\begin{abstract}		
	Millimeter wave (\texttt{mmWave}) channels in device to device (\texttt{D2D}) communication are susceptible to blockages in spite of using directional beams from multi-input multi-output (\texttt{MIMO}) antennas to compensate for high propagation loss. This motivates one to look for the presence of obstacles while forming \texttt{D2D} links among user equipments (\texttt{UEs}) which are in motion. In \texttt{D2D} communication, moving \texttt{UEs} also act as relays to forward data from one \texttt{UE} to another which introduces the problem of relay selection. The problem becomes more challenging when the obstacles are also in motion (dynamic obstacles) along with the moving \texttt{UEs}. First we have developed a probabilistic model for relay selection which considers both moving \texttt{UEs} and dynamic obstacles. Then we have analyzed the probability of dynamic obstacles blocking a link in 3D Euclidean space by exploiting the information from \texttt{MIMO} radar connected to the base station. Finally, using this information, we have developed unique strategies based on simple geometry to find the best relay which maximizes the expected data rate. Through simulations we have shown that our proposed strategy gives a significant improvement in packet loss due to mobility of nodes and dynamic obstacles in a \texttt{mmWave} channel over traditional approaches which do not consider dynamic obstacle's presence.
	\end{abstract}
	\begin{IEEEkeywords}
	\texttt{5G} \texttt{D2D} communication, Millimeter wave, Dynamic obstacles, Mobile \texttt{UEs} and Packet loss.
	\end{IEEEkeywords}
	
	\IEEEpeerreviewmaketitle

	\maketitle
	\section {Introduction}

	Device to device communication (\texttt{D2D}) in \texttt{5G} is one of the features where proximity  user equipments (\texttt{UEs}) have the potential to bypass base station (\texttt{BS}) to form \texttt{D2D} links among themselves \cite{tehrani2014device}.  Recently,  millimeter wave (\texttt{mmWave}) is widely studied for  short range \texttt{D2D}  communication  \cite{8014297} due to their high available bandwidth and capacity. Although \texttt{mmWave}  suffer from higher propagation loss characteristics, it is compensated by placing a large number of antennas in a small region owing to their  smaller wavelength, which in turn  increases the antenna gains at transmitter and receiver. These multi-input multi-output (\texttt{MIMO}) antennas make the  directional communication possible using beam-forming techniques \cite{7974772}. However, \texttt{mmWave} channels are very much susceptible to the blockage by obstacles due to very high penetration loss. For example, penetration losses of about $40 ~dB$ for outdoor tinted glass at 28 GHz \texttt{mmWave} and  $178 ~dB$ from a $10 ~cm$ brick wall at 40 GHz are mentioned in \cite{6655403} and \cite{5783993} respectively. This requires almost a line of sight (\texttt{LOS}) communication for a given \texttt{D2D} link. Note that there exist studies which consider reflected waves for short range \texttt{D2D} communication as in  \cite{p3_7842213}, but we focused on \texttt{LOS} communication in this work. 
	
	In \texttt{D2D} communication, relay selection is one of the important problem to be studied to achieve higher throughput and data rate.  Relays can be used to divert the communication path to mitigate the effects of outages due to blockages. To mitigate obstacle's severe effect, several studies exist in literature \cite{6840343,6932503,p11_7504422}. The authors in  \cite{p11_7504422}  and \cite{p8_7510705} studied  relay-assisted \texttt{mmWave} communication to alleviate blockages. Similarly \texttt{D2D} relays for \texttt{mmWave} cellular system subject to Bernoulli blockages was studied in \cite{p14_7450161} and performance of \texttt{D2D} 2-hop relays to overcome blockages using stochastic geometry was studied in \cite{p13_8241868}. However, these studies take into account the static nature of obstacles which may not be true in practice where there may be dynamic obstacles moving throughout the given service area. The problem arising due to uncertainty caused by dynamic obstacles becomes more  challenging when the nodes or \texttt{UEs} participating in the \texttt{D2D} communication are also in motion \cite{DBLP:conf/nca/distributed}. In this case, even static obstacles become dynamic relative to moving \texttt{UEs}. However, this scenario of static obstacles can be dealt with ease, but the difficulty in the problem arises when moving obstacles come into picture. 
	 
	 To account for the dynamic nature of the obstacles we have leveraged the information from radars which uses Doppler effect, a phenomenon  widely applied in various domains. In wireless communication, it has been studied to synchronize wireless sensor nodes \cite{D-sync},  to recognize human gesture \cite{Pu:2013:WGR:2500423.2500436},  to locate people \cite{Adib:2015:MLV:2789770.2789790} and recently studied in \cite{8038804} for search and rescue operations  to locate trapped people in natural disasters. High reflection coefficient of \texttt{mmWave}  in outdoor materials \cite{6655403} makes them suitable for utilizing the Doppler effect phenomenon as in radars.  Blockage detection performance of radars co-deployed with cellular system  is analyzed in \cite{8457255}. Recently, efforts have been made to leverage the \texttt{MIMO} radar along with   \texttt{mmWave} communication system \cite{8422132}. \texttt{MIMO} radar has also been the focus of recent  study to get more accurate information of position of objects \cite{7541532,7885501}. Since the motion of \texttt{UE}  along with dynamic obstacles is inevitable to analyze the probable \texttt{LOS} communication which is important criteria for choosing the best \texttt{mmWave} link. Hence, we developed a strategy to choose such a link in an environment where \texttt{UEs} and obstacles both are moving. In this paper, we study the problem of relay selection under this environment where an appropriate best relay node has to be chosen from the set of available relays for source-destination communication  by incorporating the dynamics of obstacles and \texttt{UE} motion leveraging the use of radars. We first design a simple geometric technique to capture the movement of dynamic obstacles and \texttt{UEs}. Then using this geometric analysis, we develop an algorithm to choose the best relay from the set of available relays which can provide the maximum expected data rate. Finally, we have compared and shown that our algorithm outperforms traditional approaches which do not consider the effect of dynamic obstacles.

		 \section{System Model} \label{system_model}
		 \subsection{Network}
\begin{figure}[h!]
		\centering
		\includegraphics[width=0.4\textwidth]{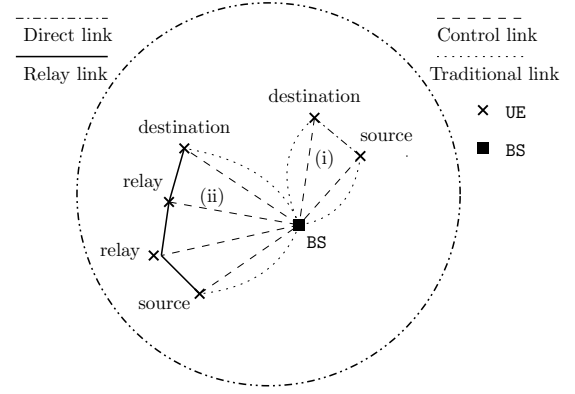}
		\caption{Network-assisted device-tier architecture for \texttt{D2D} communication \cite{tehrani2014device}  }
		\label{fig:d2d}
	\end{figure}

	 We are considering  a service region occupied with mobile nodes which can form \texttt{D2D} communication and a single base station (\texttt{BS}). We are considering specifically the operator-controlled (network-assisted) scenario of device-tier of \texttt{5G} \texttt{D2D} architecture mentioned in \cite{tehrani2014device} and shown in figure \ref{fig:d2d}, where the \texttt{BS} assists  \texttt{UEs} for either a direct connection between them or through a potential \texttt{UE} relay to forward data to the destination. Time is discretized  as $t$, $t+1$, $t+2$, $\ldots$, where $\Delta t$ is the small time difference between the current time instance $t$ and the next time instance $t+1$. At time instant $t$, the connectivity of the mobile nodes or \texttt{UEs} is represented as a graph $G^t(N^t,E^t)$, where $N^t$ represents the set of \texttt{UEs} and $E^t$ represents the set of edges. Here an edge $(i,j)$ between two \texttt{UEs} $i \in N^t$ and $j \in N^t$ represents that they can communicate to each other. For a node $i\in N^t$, $adj^t(i)$ is the set of all neighbors of node $i$ at time $t$. We assume that \texttt{UEs} are moving independently and the links are formed independently of each other. We are considering device-tier of the aforementioned \texttt{D2D} architecture, hence we are assuming that \texttt{UEs} have the capability to form \texttt{mmWave-D2D} link among themselves for \texttt{D2D} communication in out-band or in-band overlay scenarios such that they are not interfered from the cellular users. Node $i\in N^t$ is moving with velocity vector $\overrightarrow{V}_i^t$. We denote speed, angle of elevation and azimuth angle of node $i$ at time $t$ as $V_i^t$, $\alpha_i^t$ and $\beta_i^t$  respectively, which are known at the \texttt{BS}.  For each node $i\in N^t$, its  acceleration is $0$ for $\Delta t$ time duration (speed  is unchanged for $\Delta t$ duration). We are considering nodes as point objects in 3D Euclidean space. Position vector of node $i$ is defined as $\overrightarrow{T_i^{t}}:(x_i^t,y_i^t,z_i^t)$ at time $t$. Euclidean distance between nodes $i$ and $j$ moving with speeds $V_i^t$ \& $V_j^t$ respectively is denoted as $d_{ij}^{t}$.  Hence $d_{ij}^{t+1}$ can be computed as:
	 \[ d_{ij}^{t+1}=\sqrt{(z_j^{t+1}-z_i^{t+1})^2+(y_j^{t+1}-y_i^{t+1})^2+(x_j^{t+1}-x_i^{t+1})^2}\]
	 where,
	  \[x_i^{t+1}=x_i^t+V_i^{t}\Delta t\cos\alpha_i^t\sin\beta_i^{t};~ x_j^{t+1}=x_j^t+V_j^{t}\Delta t\cos\alpha_i^t\sin\beta_i^{t}\]
	  \[y_i^{t+1}=y_i^t+V_i^{t}\Delta t\cos\alpha_i^t\cos\beta_i^{t};~ y_j^{t+1}=y_j^t+V_j^{t}\Delta t\cos\alpha_i^t\cos\beta_i^{t}\] \[z_i^{t+1}=z_i^t+V_i^{t}\Delta t\cos\alpha_i^t;~z_j^{t+1}=z_j^t+V_j^{t}\Delta t\cos\alpha_i^t\]
	Note that azimuth angle is measured with respect to positive y-axis (north direction when look from top into the x-y plane) and elevation angle is measured with respect to the x-y plane (horizon) as shown in figures \ref{fig:both_model}(i)-(ii).  Elevation and azimuth angles  are  two important parameters which signify respectively height of the reflecting objects and their orientation with respect to the positive y-axis direction (north direction) on the given plane.  Azimuth angle along with round trip delay of reflected wave gives the position of object at current time instance in x-y plane. Note that \texttt{UEs} could be tracked easily as they are connected to \texttt{BS}, however it is difficult to track other dynamic objects which are not connected to \texttt{BS} like vehicles and people. We are assuming that the \texttt{BS} has \texttt{MIMO} radar capability that can be used to measure  $\alpha_i^t$, $\beta_i^t$ and round trip delay using Doppler effect. Here  $\alpha_i^t$ and $\beta_i^t$ signify the moving direction of \texttt{UEs} in 3D Euclidean space, round trip delay helps to measure the distance $r_i^t$ of node $i$ from the \texttt{BS} at time $t$ and Doppler shift measures the speed of node $i$. Note that  $r_i^t$ and $r_j^t$ can be used to measure positions $\overrightarrow{T_i^{t}}$  and $\overrightarrow{T_j^{t}}$ of nodes $i$ and $j$ respectively with respect to the \texttt{BS}. Hence we can measure distance $d_{ij}^{t}$ between nodes $i$ and $j$.
	 
	 \textit{Obstacles:} We are considering that a link $(i,j) \in E^t$ can be obstructed by some static and dynamic obstacles. The dynamic obstacles may not be communicating with the \texttt{BS}, which might bring difficulty in tracking them. Hence we are using radar leveraged \texttt{BS} which can detect the presence of obstacles with some probability of success as discussed in \cite{8457255}, where the obstacles are treated as line Boolean model with their centers distributed according to independent homogeneous Poisson point process (PPP) with density $\Lambda_o$. The length and orientation of each obstacle are uniformly distributed and written as $\eta_k$ and $\theta_k$ respectively. We denote $\mathbb{K}=\{1,2,\cdots,K\}$ as the set of dynamic obstacles which are moving independent of each other, where $K$ is the maximum number of obstacles present in the given region. There are some radars located in that area and radar locations are also independent homogeneous PPP with density $\Lambda_R$. The presence of an obstacle $k$ is detected with the closest radar with detection probability $p_k^t$ at time $t$ as mentioned in \cite{8457255}. For analysis, we are considering the center of the dynamic obstacle $k \in \mathbb{K}$ as a point object with position vector $\overrightarrow{T_k^{t}}:(x_k^t,y_k^t,z_k^t)$. Similarly there are a total of $L$ static obstacles and an static obstacle $l \in \mathbb{L}$  remains stationary throughout the experiment. Hence its position can be represented as: $\overrightarrow{T_l}:(x_l,y_l,z_l)$. Their positions can be pre-computed in a lookup table and can be easily verified for their interference with a communicating \texttt{D2D} link. Let us denote $I_{ij}^{t+1}$ as the indicator variable representing if any obstacle $k \in \mathbb{L} \cup \mathbb{K}$ blocks link $(i,j)$ under consideration when communication takes place during time interval $\Delta t$ from the current time instant $t$ to the next time instant $t+1$:
	 \begin{equation}
	 I_{ij}^{t+1}=\begin{cases}
	 0, & \text{if}  ~(i,j)~ not~ blocked ~by ~any~ k\in \mathbb{L} \cup \mathbb{K} ~at~ t+1\\
	 1, & \text{otherwise}  \\
	 \end{cases} \label{eq:blockage_indicator}
	 \end{equation}	
	\subsection{\texttt{mmWave} Channel}
	 We are considering a simple sectored antenna array model for both transmitters and receivers. For an $M\times M$ uniform planar square antenna array, antenna gains can be written as \cite{p12_7925803}: 
	\begin{equation}
	G_{x}=\begin{cases}
	G_{ml} & \text{if}\ \theta \le \phi/2 \\
	G_{sl}, & \text{otherwise}  \\
	\end{cases} \label{eq:mmw_sidelobe}
	\end{equation}
	where $x=\{t,r\}$ is subscript for transmitter \& receiver, $G_{ml}=M^2$,  $G_{sl}$ and $\phi$ are main-lobe gain, side-lobe gain and beam-width respectively. Here $\theta\in[-\pi,\pi]$ is the angle off the bore-sight direction. We are assuming that the transmitter-receiver pairs are perfectly aligned to obtain the maximum power gain \cite{7974772}. Alignment overhead is in order of  hundreds of micro seconds even for extremely narrow beams of width=$1^{\circ}$ as mentioned and validated in  \cite{7562151} which can be neglected with respect to communication time in order of seconds.
	
	For a  link $(i,j) \in E^t$, where node $i$ is the transmitter and $j$ is receiver, the received signal to noise ratio (\texttt{SNR}) can be computed as $S_{ij}^{t+1}=\frac{Q_{ij}^{t+1}}{N_{th}}$, where $N_{th}$ is the constant thermal noise, $Q_{ij}^{t+1}$ is the  signal strength received at $j$ from $i$ at time $t+1$ in presence of obstacles. We are considering both the large scale fading path loss as well as path loss due to penetration from the obstacles. Here $Q_{ij}^{t+1}$ consists of these two components and can be defined as \cite{wireless_communications,8386686}: 
	\begin{equation}
	Q_{ij}^{t+1}= \overline{pl}_{ij}^{t+1} \cdot \overline{pp}_{ij}^{t+1} \label{eq_received_signal}
	\end{equation} 
	where $\overline{pl}_{ij}^{t+1}=P_{i}^t\cdot K\cdot\bigg(\frac{d_0}{d_{ij}^{t+1}}\bigg)^\rho\cdot\psi$  is a component of the received power due to fading of signal and $\overline{pp}_{ij}^{t+1}=\frac{1}{\Gamma_p}$ is the component of received power due to penetration loss because of obstacle's presence in link $(i,j)$. Here $P_{i}^t$ is the transmitted power from  node $i$ at time $t$ and $K=G_t\cdot G_r\cdot \big(\frac{\lambda}{4\pi d_0}\big)^2$ is a constant. Here $G_t$ and $G_r$ are the transmitter  \& receiver antenna gains respectively which are assumed to be constant over time,  $\lambda$ is the wavelength and $d_0$  is a reference distance for the antenna far-field. Here $d^{t+1}_{ij}$ is the distance between nodes $i$ and $j$ at time $t+1$, $\rho$ is the path loss exponent (\texttt{PLE}) and $\psi$ is the shadowing random variable. We also assume $P_i^t$ to be constant for each transmitting node. $\Gamma_p$ is the penetration loss from the blocking obstacle. We are assuming that the penetration loss by a single obstacle $\Gamma_p\rightarrow\infty$ \cite{7562151} and thus  even a presence of single obstacle may break the connectivity of the given \texttt{mmWave} link. Later, $I_{ij}^{t+1}$ is computed by making use of this fact. This implies an  \texttt{LOS} path is required between two \texttt{D2D} nodes for successful communication. 
	\section{Problem Formulation \& Probabilistic Model} \label{problem_formulation}
Suppose a mobile node $i$ transmits at time $t$ to another mobile node $j$ (relay or destination node) by forming a link $(i,j)$. The transmission takes place for  $\Delta t$ duration till the next  time instance $t+1$. The link may be disconnected due to mobility of the nodes or may be blocked by some obstacle during this $\Delta t$ duration. Let us define $e^{t}_{ij}$ as a Boolean variable whose value is $1$, if there is an edge between nodes $i$ and $j$ at time $t$, and $0$, otherwise. Here an edge between two nodes represents that they are within the communication range of each other and there is no obstacle between them. Our problem is to find out those links which are connected at current time instant $t$ (i.e., $e^{t}_{ij}=1$) and have the higher probability of being connected for the next time instant $t+1$ while the communication takes place. Our objective is to maximize the expected data-rate while taking care of packet loss and average delay. For a given node $i$, we want to find a node $j \in adj^t(i)$ for relaying the packets such that the following objective is satisfied: 
	\begin{equation}
		\argmax_j E[C_{ij}^{t+1}],  j \in adj^t(i)
	\end{equation} 
	where, $E[\cdot]$ denotes the expectation, $C_{ij}^{t+1}$ denotes the  capacity  of link $(i,j)$ at the next time instance $t+1$. Thus $E[C_{ij}^{t+1}]$ signifies the expected data rate available till next time instance $t+1$. Let us define $\mathbb{S}^{t+1}_{ij}$ as the probability that link $(i,j)$ will be connected at time $t+1$ given it was connected at current time instance $t$:
	\begin{equation} \label{eq:condP}
	\mathbb{S}_{ij}^{t+1}  = P\{e_{ij}^{t+1}=1|e_{ij}^t=1\},  ~\forall(i,j) \in E^t. 
	\end{equation}
	Now, we can write $E[C_{ij}^{t+1}]=\mathbb{S}_{ij}^{t+1}\cdot C_{ij}^{t}+(1-\mathbb{S}_{ij}^{t+1})\cdot C_{ij}^{t}$.	We know that with probability $(1-\mathbb{S}_{ij}^{t+1})$, link $(i,j)$ is going to fail at the next time instance $t+1$, hence $C_{ij}^{t+1}=0$. Hence, our objective reduces to:
	\begin{equation} \label{objective_final}
		E[C_{ij}^{t+1}]=\mathbb{S}_{ij}^{t+1}\cdot C_{ij}^{t}.
	\end{equation}
	$\mathbb{S}^{t+1}_{ij}$ captures the link breakage probability during the transmission time $\Delta t$  considering nodes mobility as well as static and dynamic obstacles for the upcoming time instance $t+1$. Note that  $\mathbb{S}^{t+1}_{ij}$ is computed at  current time instance $t$. It is evident that $e_{ij}^{t+1}$ and $e_{ij}^t$ are independent as nodes $i$ and $j$ are moving independently and also  $K$ obstacles are moving independently of each other. The independence of $e_{ij}^{t+1}$ and $e_{ij}^t$ also arises from the fact that even for two static nodes $i$ and $j$ which are connected at time $t$, may get disconnected at upcoming time instance $t+1$ due to blockage from independently moving obstacles. Hence we may reduce equation \eqref{eq:condP} to:
	 \begin{equation} \label{stoc_constr}
	 \mathbb{S}_{ij}^{t+1}  = P\{e_{ij}^{t+1}=1\}. 
	 \end{equation}

	To satisfy equation \eqref{stoc_constr}, the \texttt{SNR} received at node $j$ from node $i$ must be greater than a minimum threshold  $S_{ij}^{th}$. The threshold $S_{ij}^{th}$ denotes the required \texttt{SNR} of the given link $(i,j)$ depending upon the type of communication used (e.g., voice, video call etc.). So equation \eqref{stoc_constr} reduces to: 
 \begin{equation} \label{stoc_constr_new}
\mathbb{S}_{ij}^{t+1}=P(S_{ij}^{t+1}\ge S_{ij}^{th}).
\end{equation}

Since $S_{ij}^{t+1}$ is a function of $Q_{ij}^{t+1}$, for the upcoming time instance $t+1$, $S_{ij}^{t+1}$ depends on (i)  $\overline{pl_{ij}^{t+1}}$ and (ii) $\overline{pp_{ij}^{t+1}}$. Now we can compute $\overline{pp_{ij}^{t+1}}$ using the indicator random variable $I_{ij}^{t+1}$ as stated in equation \eqref{eq:blockage_indicator}. Hence we can express equation \eqref{stoc_constr_new} as a joint distribution of $\overline{pl_{ij}^{t+1}}$  and  $I_{ij}^{t+1}$:
	\begin{equation}
	\mathbb{S}_{ij}^{t+1}=P(\overline{pl_{ij}^{t+1}}\ge\gamma_{ij},I_{ij}^{t+1}=0)\label{eq_distance_function}
	\end{equation}	
	where $\gamma_{ij}$ is the threshold on received power to satisfy the given data-rate requirements and $I_{ij}^{t+1}$ indicates that the link is not blocked in the upcoming time instant $t+1$. We can write equation \eqref{eq_distance_function} as a conditional probability expression:
	\begin{equation}
	\mathbb{S}_{ij}^{t+1}=P(\overline{pl_{ij}^{t+1}}\ge\gamma_{ij} | I_{ij}^{t+1}=0 )\cdot P(I_{ij}^{t+1}=0)\label{eq_independent}
	\end{equation}
	The first term of right hand side in above equation signifies the probability of packet loss  due to node's mobility  when there is no obstacle  and the second term takes care of  probability that  whether any obstacle  interferes with the given link till the next time instance $t+1$. Now in subsequent sections we will show how to compute these two terms.
	 \begin{figure*}[h!]
		\centering
		\includegraphics[width=0.9\textwidth]{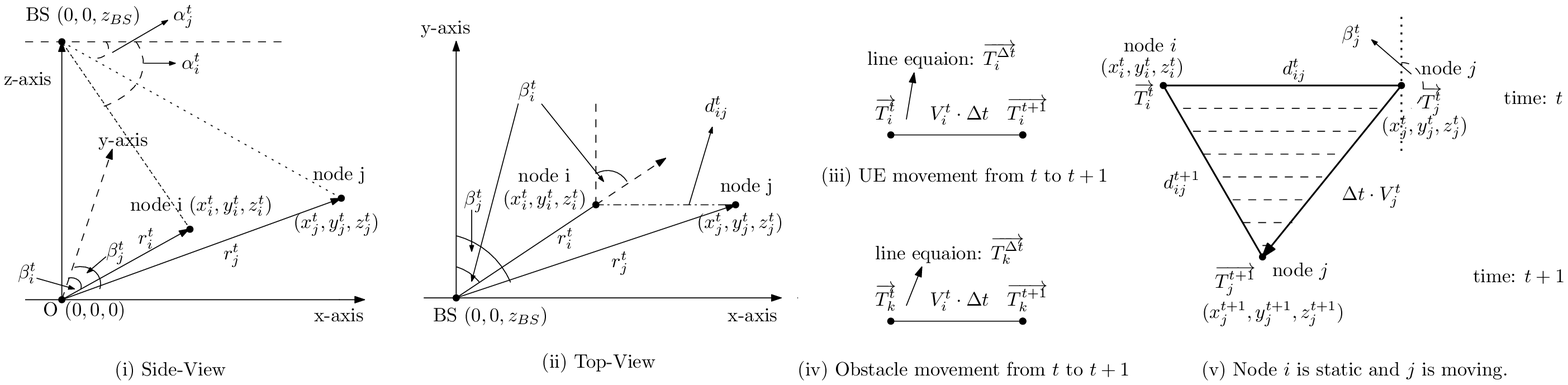}
		\caption{Position, orientation and representation of path of movement for \texttt{UEs} and dynamic obstacle.  }
		\label{fig:both_model}
	\end{figure*}
	 \section{Analysing Effects of Mobility and Obstacles in Relay Selection} \label{doppler}
	  We are exploiting the information from the radar linked with the \texttt{BS} to locate the \texttt{UEs} and moving obstacles. The positions are first find out and $P(\overline{pl_{ij}^{t+1}}\ge\gamma_{ij} | I_{ij}^{t+1}=0 )$ is computed based on it. Then using geometrical analysis, we compute $P(I_{ij}^{t+1}=0)$.  Finally, we  develop an algorithm using the analysis to determine the best relay node among the potential relays considering both static and dynamic obstacles. 
	 \subsection{Finding Positions and Movements of Nodes and Obstacles}
	 \texttt{BS} will store the location of moving \texttt{UEs} and obstacles using the analysis as shown in figures \ref{fig:both_model}(i)-(ii). Figure \ref{fig:both_model}(i) shows the side-view and figure \ref{fig:both_model}(ii) shows its top-view indicating distance $d_{ij}^t$ between nodes $i$ and $j$.  \texttt{BS} of height $z_{BS}$ is located at $(0,0,z_{BS})$, nodes $i$ \& $j$ are located at $(x_i^t,y_i^t,z_i^t)$ \& $(x_j^t,y_j^t,z_j^t)$ at time $t$ respectively. We can find out the positions of node $i$:
	  \[x_i^t=r_i^t\cos\alpha_i^t\sin\beta_i^t;y_i^t=r_i^t\cos\alpha_i^t\cos\beta_i^t;z_i^t=z_{BS}-r_i^t\sin\alpha_i^t\]
Similarly we can compute the positions of static and dynamic obstacles.	 Once the respective positions are known, we need to analyze if the links formed at time $t$ are going to be obstructed by any obstacle for $\Delta t$ time duration till the time instance $t+1$ or not. To do so, we need to look into the path of a moving object (\texttt{UE} or dynamic obstacle) for $\Delta t$ duration as follows: a moving \texttt{UE}  $i$  positioned at $\overrightarrow{T_i^{t}}$ at time $t$ will move with velocity $V_i^t$ for duration of $\Delta t$ to arrive at new location $\overrightarrow{T_i^{t+1}}$ at time $t+1$.  This movement for a short time duration $\Delta t$ is a straight line as shown in figure \ref{fig:both_model}(iii). The equation of this line segment is:
	 \begin{equation}
	 	\delta\cdot\overrightarrow{T_i^{t+1}}+(1-\delta)\cdot\overrightarrow{T_i^{t}}=\overrightarrow{T_i^{\Delta t}} \label{eq_line_segment}
	 \end{equation}
	 where, $\delta\in(0,1)$. Similarly the motion path of a dynamic obstacle $k \in \mathbb{K}$ as shown in figure \ref{fig:both_model}(iv) can be written as: 
	  \begin{equation}
	 \delta\cdot\overrightarrow{T_k^{t+1}}+(1-\delta)\cdot\overrightarrow{T_k^{t}}=\overrightarrow{T_k^{\Delta t}} \label{eq_obstacle_line_segment}
	 \end{equation}
	 where, $\delta\in(0,1)$. Static obstacle $l\in\mathbb{L}$ is assumed to be positioned as $\overrightarrow{T_l}$ which is stationary throughout the experiment. 
	 
	 We can compute $d_{ij}^{t+1}$  using  $\overrightarrow{T_i^{t+1}}$ and $\overrightarrow{T_j^{t+1}}$. Then using this value, we can compute  $\overline{pl_{ij}^{t+1}}$ for a given realization of $\psi$ and respective \texttt{PLE}s. Hence $P(\overline{pl_{ij}^{t+1}}\ge \gamma_{ij} | I_{ij}^{t+1}=0 )$ can be computed as:
	 \begin{equation}
	 P(\overline{pl_{ij}^{t+1}}\ge \gamma_{ij} | I_{ij}^{t+1}=0 )=\begin{cases}
	 1, & \text{if}\  \overline{pl_{ij}^{t+1}} \ge \gamma_{ij} \\
	 0, & \text{otherwise}.  \\
	 \end{cases} \label{eq:distance_next}
	 \end{equation}
	 It signifies the minimum received signal criteria between \texttt{UEs} to satisfy the respective minimum data rate requirement of  link $(i,j)$ given that there are no interfering obstacles. In the next section,  we will compute $P(I_{ij}^{t+1}=0)$.
	 \subsection{Analyzing Blockage Due to Obstacles}
	We must capture the location of static obstacles and motion path  of dynamic obstacles in order to find $P(I_{ij}^{t+1}=0)$ as:
	 \begin{equation}
	P(I_{ij}^{t+1}=0)=\prod_{l\in\mathbb{L}}\prod_{k\in\mathbb{K}}p_k^t\cdot P_{ij\_k}^{int} \cdot P_{ij\_l}^{int} . \label{eq:I_0}
	\end{equation}
	where $p_k^t$ is the detection  probability of dynamic obstacle $k\in \mathbb{K}$ at time $t$. $P_{ij\_k}^{int}$ and $P_{ij\_l}^{int}$ are the probabilities that the link under consideration is not blocked by any of the obstacles in the set $\mathbb{K}$ and $\mathbb{L}$ respectively. Here $p_k^t$ can be computed as stated in section \ref{system_model}.	We will now calculate $P_{ij\_k}^{int}$  and $P_{ij\_l}^{int}$ for various possible cases. 
	 \subsubsection{Both nodes are stationary} \label{sec-1}
In this case, for nodes $i$ and $j$, $\overrightarrow{T_i^{t}}=\overrightarrow{T_i^{t+1}}$ and $\overrightarrow{T_j^{t}}=\overrightarrow{T_j^{t+1}}$ and hence the equation  of line $\overrightarrow{T_{ij}}$ connecting them can be expressed as:	 
	 \begin{equation}
		 \delta\cdot\overrightarrow{T_i^{t}}+(1-\delta)\cdot\overrightarrow{T_j^{t}}=\overrightarrow{T_{ij}},~ \forall \delta\in(0,1). \label{eq:stationary_line_eq} 
	 \end{equation}
	 Now we need to capture the potential obstacles which might hinder the communication between nodes $i$ and $j$ positioned at $\overrightarrow{T_i^{t}}$ and $\overrightarrow{T_j^{t}}$.  If none of the static obstacles $l\in \mathbb{L}$ positioned at $\overrightarrow{T_l}$ satisfy equation \eqref{eq:stationary_line_eq}, then $ P_{ij\_l}^{int}=1$ otherwise $ P_{ij\_l}^{int}=0$. For all dynamic obstacles $k\in \mathbb{K}$, we have to find whether the equation of their motion path $\overrightarrow{T_k^{\Delta t}}$ (equation \eqref{eq_obstacle_line_segment}) intersects with $\overrightarrow{T_{ij}}$ (equation \eqref{eq:stationary_line_eq}). Hence we can find,
	 \begin{equation}
	 P_{ij\_k}^{int}=\begin{cases}
	 1, & \text{if}\  \overrightarrow{T_k^{\Delta t}} ~do~ not ~intersect~ \overrightarrow{T_{ij}}, \forall k \in \mathbb{K} \\
	 0, & \text{otherwise}  \\
	 \end{cases} \label{eq:obstacle_pr_1}
	 \end{equation}
	 \begin{figure}[h!]
	 	\centering
	 	\includegraphics[width=0.45\textwidth]{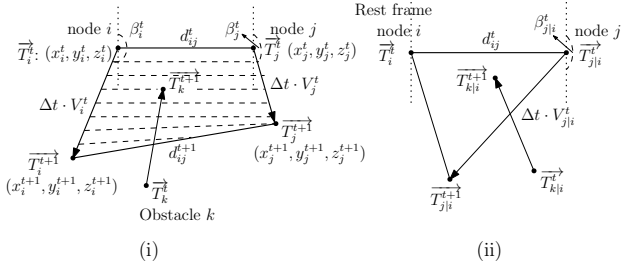}
	 	\caption{(i) Representation of both nodes  moving in a skew path and obstacle $k$ (ii) Relative motion of node $j$ and obstacle $k$ relative to node $i$.}
	 	\label{fig_stationary_moving_case}
	 \end{figure}
	 \subsubsection{One of the two nodes is moving} \label{sec-2}	 
	 Let us assume that node $i$ is stationary node and $j$ is the moving  node. We can categorize it into two cases: first is a special case when the \texttt{BS} detects that node $j$ is moving  towards or away from  the stationary node $i$ where the angle of movement  is $180^{\circ}$ or $0^{\circ}$ respectively with respect to line $\overrightarrow{T_i^{t}}\overrightarrow{T_j^{t}}$. In this case, node $j$'s movement forms a straight line with respect to the stationary node $i$ for duration $\Delta t$ denoted as $\overrightarrow{T_{ij}^{\Delta t}}=\overrightarrow{T_i^{t}}\overrightarrow{T_j^{t+1}}$. For static obstacles again if none of $l \in \mathbb{L}$ satisfy line $\overrightarrow{T_{ij}^{\Delta t}}$,  then $ P_{ij\_l}^{int}=1$ otherwise $ P_{ij\_l}^{int}=0$. For dynamic obstacles, we need to find out if any obstacle $k \in \mathbb{K}$ blocks the communication between nodes $i$ and $j$ by verifying if $k$'s motion path equation $\overrightarrow{T_{k}^{\Delta t}}$ intersects $\overrightarrow{T_{ij}^{\Delta t}}$ or not. Thus for dynamic obstacles we can compute:
	\begin{equation}
	P_{ij\_k}^{int}=\begin{cases}
	1, & \text{if}\  \overrightarrow{T_k^{\Delta t}}~do~ not ~intersect~ \overrightarrow{T_{ij}^{\Delta t}} , \forall k \in \mathbb{K} \\
	0, & \text{otherwise}  \\
	\end{cases} \label{eq:obstacle_pr_4}
	\end{equation}
	
	The second case is described when node $j$ moves at angle relative to node $i$ other than that from the set $\{0^{\circ},180^{\circ}\}$. This case is described in figure \ref{fig:both_model}(v). Initially at time $t$, node $j$ is inside the range of node $i$ at a distance of $d_{ij}^t$ when no obstacles were present.  By the next time instance $t+1$, node $j$ will cover a distance of $\Delta t\cdot V_j^t$ from its initial point $\overrightarrow{T_j^{t}}$. As mentioned in equation \eqref{eq_line_segment}, $\overrightarrow{T_j^{\Delta t}}=\overrightarrow{T_j^{t}}\overrightarrow{T_j^{t+1}}$ denotes the line segment representing the movement of node $j$ for the duration of $\Delta t$ time.
	In the mentioned figure, arrow  denotes the direction of motion of node $j$. The shaded region denotes a bounded region $B_{ij}^{\Delta t}$ formed by three vertices $\overrightarrow{T_i^t}$, $\overrightarrow{T_j^t}$ and $\overrightarrow{T_j^{t+1}}$ during time $\Delta t$. Since three points define a unique plane in 3D Euclidean space, points in $B_{ij}^{\Delta t}$ are coplanar and the equation of the plane denoted as $P_1^{\Delta t}$ is:
	 \begin{equation}
	 	(\overrightarrow{T_p^{\Delta t}}-\overrightarrow{T_i^t})\cdot((\overrightarrow{T_j^{t+1}}-\overrightarrow{T_i^t})\times (\overrightarrow{T_j^t}-\overrightarrow{T_i^t}))=0 \label{eq_plane}
	 \end{equation}
	 where $\times$ denotes vector cross product, $\overrightarrow{T_p^{\Delta t}}$ denotes the position vector $(x^t,y^t,z^t)$. The shaded region covers the entire possible area where communication between nodes $i$ and $j$ takes place. We also call this region $B_{ij}^{\Delta t}$ as the \textit{communication region} which might be vulnerable due to presence of obstacles. For  static obstacle $l, ~\forall l \in \mathbb{L}$, we  need to check if point $\overrightarrow{T_l}$  satisfies equation  \eqref{eq_plane} or not. If it does not satisfy then there is no blockage from it. Otherwise, we need to check if it lies inside the bounded region $B_{ij}^{\Delta t}$ or not. If it lies inside then $P_{ij\_l}^{int}=1$, otherwise $P_{ij\_l}^{int}=0$. For each dynamic obstacle $k\in \mathbb{K}$, we need to check whether the line segment $\overrightarrow{T_k^{\Delta t}}$ intersects with the given plane $P_1^{\Delta t }$ in equation \eqref{eq_plane}. For this, we need to consider three possible cases:
	 \begin{enumerate}[\text{case} i. ]
	 	\item $\overrightarrow{T_k^{\Delta t}}$ does not intersect with plane $P_1^{\Delta t}$ and in this case is parallel to plane.
	 	\item $\overrightarrow{T_k^{\Delta t}}$ lies on plane $P_1^{\Delta t}$.
	    \item $\overrightarrow{T_k^{\Delta t}}$ intersects plane $P_1^{\Delta t}$ on a point. In this case it crosses the plane.
 	 \end{enumerate}
	 Let us say $\hat{k}^{\Delta t}$ represents the unit vector (direction vector) of the line representing the obstacle's movement ($\overrightarrow{T_k^{\Delta t}})$ and $\overrightarrow{u}^{\Delta t}=(\overrightarrow{T_j^{t+1}}-\overrightarrow{T_i^t})\times (\overrightarrow{T_j^t}-\overrightarrow{T_i^t})$ denotes the normal vector to  plane  $P_1^{\Delta t}$. To categorize  all the above mentioned cases, we need to find the dot product of $\hat{k}^{\Delta t}$ and $\overrightarrow{u}^{\Delta t}$, if it is $0$ then $\overrightarrow{u}^{\Delta t}$ and $\hat{k}^{\Delta t}$ are orthogonal and hence the plane is parallel to $\overrightarrow{T_k^{\Delta t}}$. In this case there are two possibilities, first $\overrightarrow{T_k^{\Delta t}}$ may lie outside plane and is parallel to the plane (case i) and second when $\overrightarrow{T_k^{\Delta t}}$ lies on the plane (case ii). To further distinguish between these two, we need to find dot product of $(\overrightarrow{T_i^t}-\overrightarrow{T_k^t})$ and $\overrightarrow{u}^{\Delta t}$. If this value is $0$ then line is contained in the plane (case ii) otherwise line is outside the plane (case i). For case i,  when line $\overrightarrow{T_k^{\Delta t}}$ is outside the plane and parallel to it then obstacle $k$ does not interfere with the communication region. For case ii,  $\overrightarrow{T_k^{\Delta t}}$ lies inside  plane $P_1^{\Delta t}$ and hence has the possibility of interfering with the bounded region (which is the communication region). Now in this case, we need to figure out if obstacle's line equation $\overrightarrow{T_k^{\Delta t}}$ lies inside this region or not. To check this we perform the following two step procedure for all the dynamic obstacles $k \in \mathbb{K}$:
	 \begin{enumerate}[step a.]
	 	\item Check if moving obstacle's line equation $\overrightarrow{T_k^{\Delta t}}$ intersects with any of the three sides of the bounded region $B_1^{\Delta t}$, i.e., $\overrightarrow{T_i^{t}}\overrightarrow{T_j^{t}}$, $\overrightarrow{T_i^{t}}\overrightarrow{T_j^{t+1}}$ or $\overrightarrow{T_j^{t}}\overrightarrow{T_j^{t+1}}$ \label{step-1}.
	 	\item If step \ref{step-1} is successful, it implies obstacle $k$ interferes with $B_{ij}^{\Delta t}$. Otherwise, we need to check if line $\overrightarrow{T_k^{\Delta t}}$ is either completely inside or outside $B_{ij}^{\Delta t}$. For this, we will check for any one of the points either $\overrightarrow{T_k^{t}}$ or $\overrightarrow{T_k^{t+1}}$ of motion path of dynamic obstacle $k$  lies inside or outside $B_{ij}^{\Delta t}$. If that point lies inside then whole line segment describing the motion path of dynamic obstacle lies inside $B_{ij}^{\Delta t}$ otherwise it lies outside. If it lies inside then it interferes with the communication region during $\Delta t$ time, otherwise not. 
	 \end{enumerate}
  	For case iii, if the dot product $\hat{k}^{\Delta t}\cdot \overrightarrow{u}^{\Delta t}$ is not $0$ then line $\overrightarrow{k}^{\Delta t}$ crosses  plane $P_1^{\Delta t}$ and  they  intersect at one point (case iii). We need to check if this point of intersection lies inside bounded region $B_{ij}^{\Delta t}$. If so then obstacle $k$  interferes the communication region of nodes $i$ and $j$, otherwise there is no interference. Based on above discussion, we can calculate  $P(I_{ij}^{t+1}=0)$ as follows:	 
	 \begin{equation}
	  P_{ij\_k}^{int}=\begin{cases}
	 	1, & \text{if}\  \overrightarrow{T_k^{\Delta t}}~do ~not ~interfere~ B_{ij}^{\Delta t} , \forall k \in \mathbb{K} \\
	 	0, & \text{otherwise}  \\
	 \end{cases} \label{eq:obstacle_pr_2}
 \end{equation} 
	 \subsubsection{Both nodes are moving} \label{sec-3} 
	 In this case both nodes $i$ and $j$ are moving  from time $t$ to $t+1$. Based on their relative angle of motion we can categorize them into two cases, first when they both are moving towards or away from each other (relative angle of movement $\in\{0^{\circ},180^{\circ}\}$).  This case is similar to that of the previous section except here node $i$ is also moving. But here too, both nodes will form a straight line and hence we need to simply check if any obstacle is intersecting it. Hence this is solved in similar way as mentioned in the previous section. 
	 
	 For the second case when nodes $i$  and $j$ are \textit{not} moving towards or away from each other, there are two possibilities: case i) all four points are co-planar and case ii) they are not co-planar. Case i) is formed when either both nodes $i$ \& $j$ are moving parallel to each other or when they are intersecting each other's motion path. The path formed by them is not co-planar when the respective equations of their motion paths are skew (i.e., they neither intersect nor are parallel to each other) as shown in figure \ref{fig_stationary_moving_case}(i). To check for this categorization, we need to form a plane equation with any three points out of the given four points and then check if the fourth point lies inside this plane equation or not. Now to check whether an obstacle interferes with the communication region, we will first give a solution  for the non co-planar case and then generalize it to the co-planar case.
	 
	For case ii), where motions paths $\overrightarrow{T_i^{\Delta t}}$ and $\overrightarrow{T_j^{\Delta t}}$ are skew, we compute the relative position and  velocity of  node $j$ with reference to the other node $i$ which is kept at rest. Relative position is computed as $\overrightarrow{T}_{j|i}^t=\overrightarrow{T}_{j}^t-\overrightarrow{T}_{i}^t$ and relative velocity is computed as $\overrightarrow{V}_{j|i}^t=\overrightarrow{V}_j^t-\overrightarrow{V}_i^t$, where the magnitude speed is denoted as $V_{j|i}^t$. This is shown in figure \ref{fig_stationary_moving_case}(ii) which is reduced from figure \ref{fig_stationary_moving_case}(i), where node $j$ is moving with reference to the fixed node $i$. This in turn gives three points $\overrightarrow{T}_{i|i}^t$, $\overrightarrow{T}_{j|i}^t$ and $\overrightarrow{T}_{j|i}^{t+1}$ from which we can form the equation of a unique plane which will give a new communication region bounded by these three points. We will also compute the relative positions of all static and dynamic obstacles and relative velocities of dynamic obstacles with respect to the fixed node $i$. Now this reduces to the problem of verifying if  any of these obstacles are interfering with the communication region. This verification can be done exactly as explained in the previous section.
	 
Using same approach we can proceed for case i) when the four points are co-planar. The only difference that arises here is that the resulting relative positions lie in the same plane which contains all four points. Whereas, for the non co-planar case the resulting relative positions might shift the plane according to vector difference of their velocities and positions. Using these analysis, we now give the relay selection algorithm.  
\subsection{Relay Selection Algorithm}
Using the above analysis, we present our dynamic-obstacle (\texttt{D-Obs}) based relay selection algorithm in algorithm \ref{algo_1}. For a given sending node $i$, we will choose the relay among the nodes in $adj^t(i)$ which gives the best expected data rate. The probability that a link  connected at current time instance $t$ is still connected for the next time instance $t+1$ is calculated from the analysis section which considers all the possible cases of movements of \texttt{UEs} and moving obstacles as done in line \ref{line5}. From line \ref{line6}-\ref{line8} we get the relay node (denoted as $chosen\_j$) with the best average data-rate. Function $begin\_transmission(i,chosen\_j)$ transmits the data from node  $i$  to the chosen relay node $chosen\_j$. This process repeats for all sending nodes $i\in N^t$. Line \ref{line5} takes $O(L+K)$ computation time for a pair of sending and relaying nodes, where $L$ and $K$ are number of static and dynamic obstacles. For a given sending node $i$, to choose the best relay node, our algorithm takes $O(n(L+K))$ running time, where $n$ is the number of adjacent nodes to node $i$.
\begin{algorithm}
	\SetKwData{Left}{left}\SetKwData{This}{this}\SetKwData{Up}{up}
	\SetKwFunction{Union}{Union}\SetKwFunction{FindCompress}{FindCompress}
	\SetKwInOut{Input}{input}\SetKwInOut{Output}{output}
	\Input{$G(N^t,E^t)$,$\gamma_{ij}$, $\mathbb{K}$, $\mathbb{L}$}
	\BlankLine
	\For{$\forall i \in N^t$}{
		\If{$i$ is sending node}{
			$max\_j=0$;\\
			\For{$\forall j \in (adj^t(i))$}{
				$temp=P(\overline{pl_{ij}^{t+1}}\ge\gamma_{ij} |I_{ij}^{t+1}=0 )\cdot P(I_{ij}^{t+1}=0)\cdot C_{ij}^t$;\\ \label{line5}
				\If{$max\_j\le temp$}{ \label{line6}
					$max\_j=temp$;\\
					$chosen\_j=j$; \label{line8}
				}
			}
		$begin\_transmission(i,chosen\_j)$\\
		}
	}
	\caption{D-Obs Algorithm }\label{algo_1}
\end{algorithm}
\section{Experiment And Results}  \label{simulations}
	 {\it Simulation Environment}: 
	 We are initially distributing $30$ \texttt{UEs} uniformly in a $200~m \times 200~m$ square area. Throughout the experiment these nodes remain within the service region. For simulation purpose, we assumed that \texttt{UEs} and  obstacles are placed on the ground (i.e. x-y plane) ignoring z-axis.  Each node is moving with speed uniformly in range  $[0,V_{max}]$ m/s, for $V_{max} \in \{5,10,15,20\}$ and angle in range $[-\pi, \pi]$. In the experiment,  $\Delta t$ which is measured in  seconds, takes value from the set $\{0.5,1,1.5,2\}$. Nodes are using directional transmitter and receiver antennas for $60~GHz$ frequency with $M=4$, such that $G_r=G_t=6~dB$ and we are considering a scenario where \texttt{LOS PLE} is $2.5$ and zero mean  log-normal shadowing random variable with standard deviation $3.5$ \cite{7974772,7109864}. Thermal noise density is $-174~dBm/Hz$ \cite{6881177} and devices are using $18~dBm$ transmit power.  Capacity of each link $(i,j)$ at time $t$ is $C_{ij}^{t}=B\cdot\log_2(1+S_{ij}^t)~bits/sec$, where $B=20~MHz$ is bandwidth and \texttt{SNR} threshold $S_{ij}^{th}$ is taken to be $20$ dB \cite{8345639}. We are assuming  fixed packet  length of $65535~bytes$. There are static and dynamic obstacles initially distributed uniformly in the environment. Number of static obstacles $L=10$ is fixed throughout the experiment, whereas number of dynamic obstacles $K$ varies in range $\{0,10,20,30\}$. For all of the cases, dynamic obstacles are moving with a speed uniformly distributed in range $[0,10]~m/s$. Radars are deployed in the region with $\Lambda_R=0.001$ \cite{8457255}. We assume a single source-destination pair for simplicity and all other devices may act as relay.
	 
We run our experiments for upto $3$-hops and averaging it per-hop. We are analyzing the effect on throughput due to $K$, network load, $V_{max}$ and $\Delta t$. We are comparing the results of our algorithm with metrics based on \texttt{RSS} and a contention based forwarding (\texttt{CBF}) approach \cite{Füßler03contention-basedforwarding} which select relay node based on signal strength and shortest distance from destination respectively.  
	\begin{figure*}[h!]
		\centering
		\begin{minipage}[b]{.24\textwidth}
			\includegraphics[width=1\textwidth]{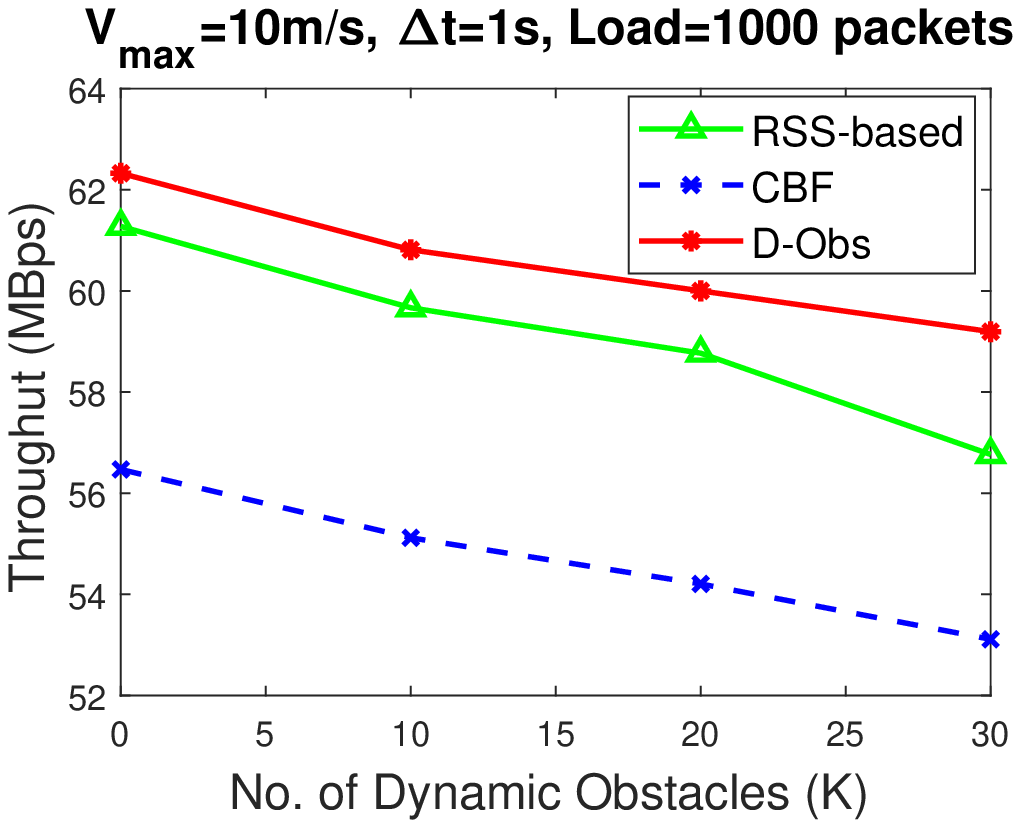}
			\caption{$K$ vs  avg. throughput. }
			\label{fig:result1}
		\end{minipage}\hspace{0.5em}
		\begin{minipage}[b]{.24\textwidth}
			\includegraphics[width=1\textwidth]{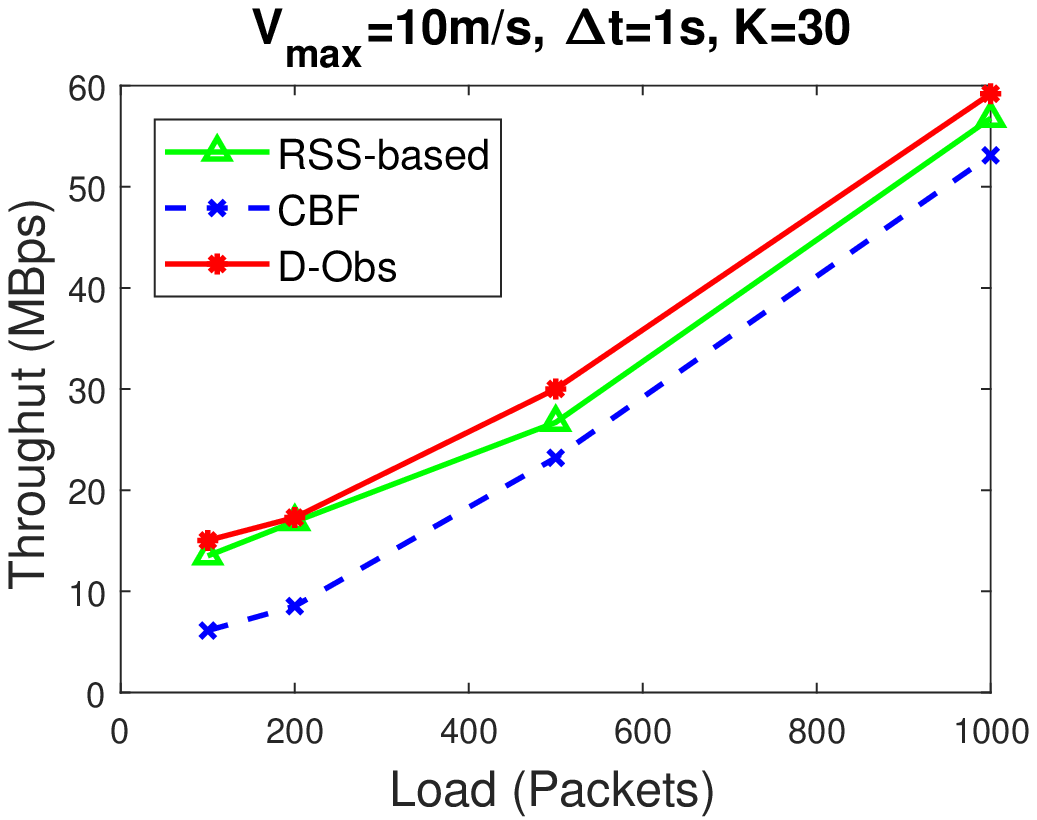}
			\caption{Load vs avg. throughput. }
			\label{fig:result2}
		\end{minipage}\hspace{0.5em}
		\begin{minipage}[b]{.24\textwidth}
			\includegraphics[width=1\textwidth]{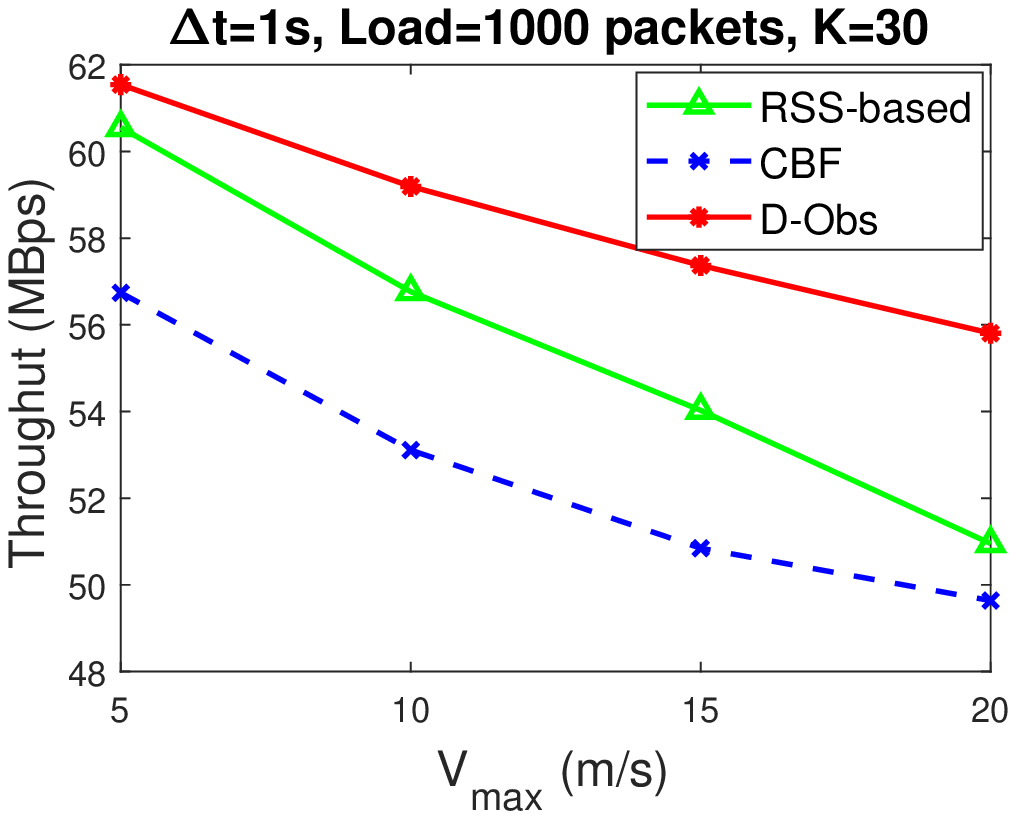}
			\caption{$V_{max}$ vs  avg. throughput.  }
			\label{fig:result3}
		\end{minipage}\hspace{0.5em}
		\begin{minipage}[b]{.24\textwidth}
		\includegraphics[width=1\textwidth]{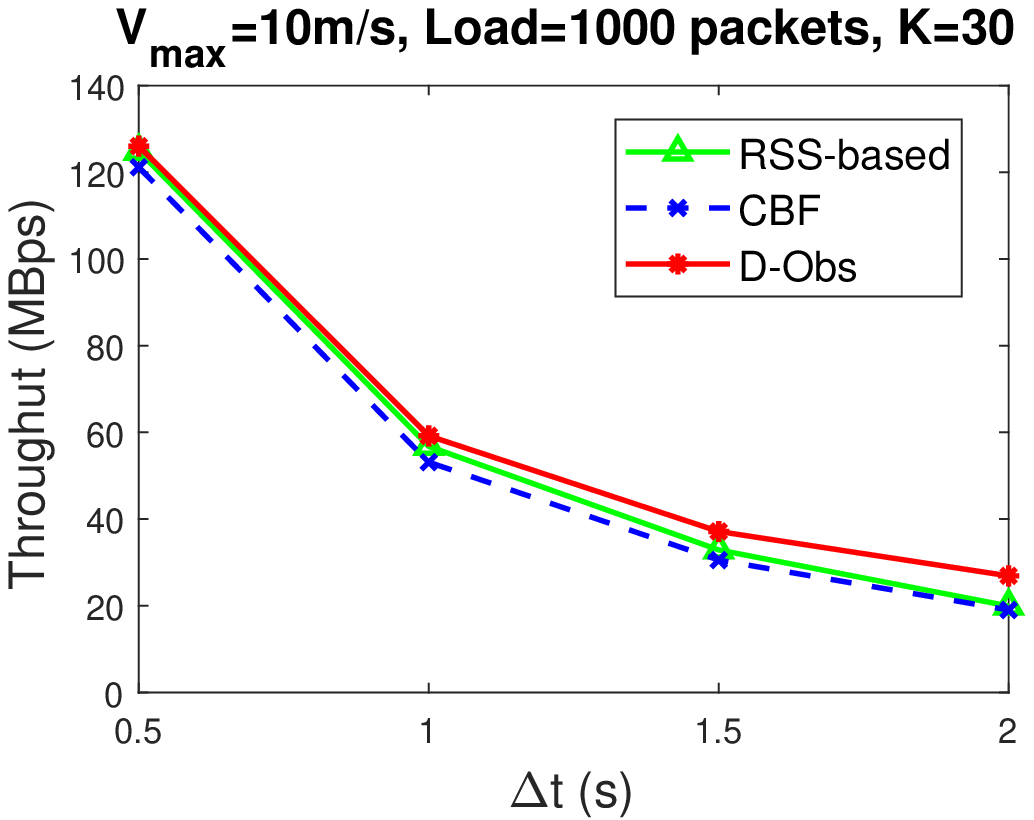}
			\caption{$\Delta t$ vs avg. throughput. }
		\label{fig:result7}
	\end{minipage}
	\end{figure*}
\begin{figure*}
		\begin{minipage}[b]{.24\textwidth}
			\includegraphics[width=1\textwidth]{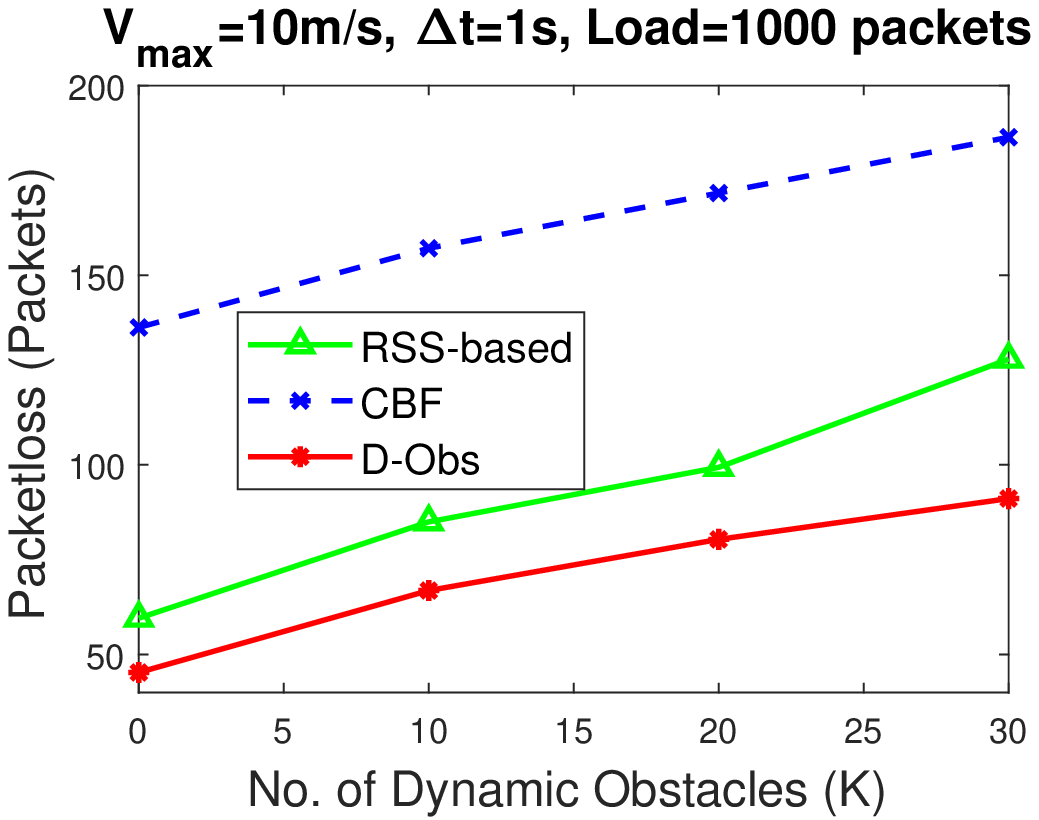}
			\caption{$K$ vs  avg. packetloss. }
			\label{fig:result4}
		\end{minipage}\hspace{0.5em}
		\begin{minipage}[b]{.24\textwidth}
			\includegraphics[width=1\textwidth]{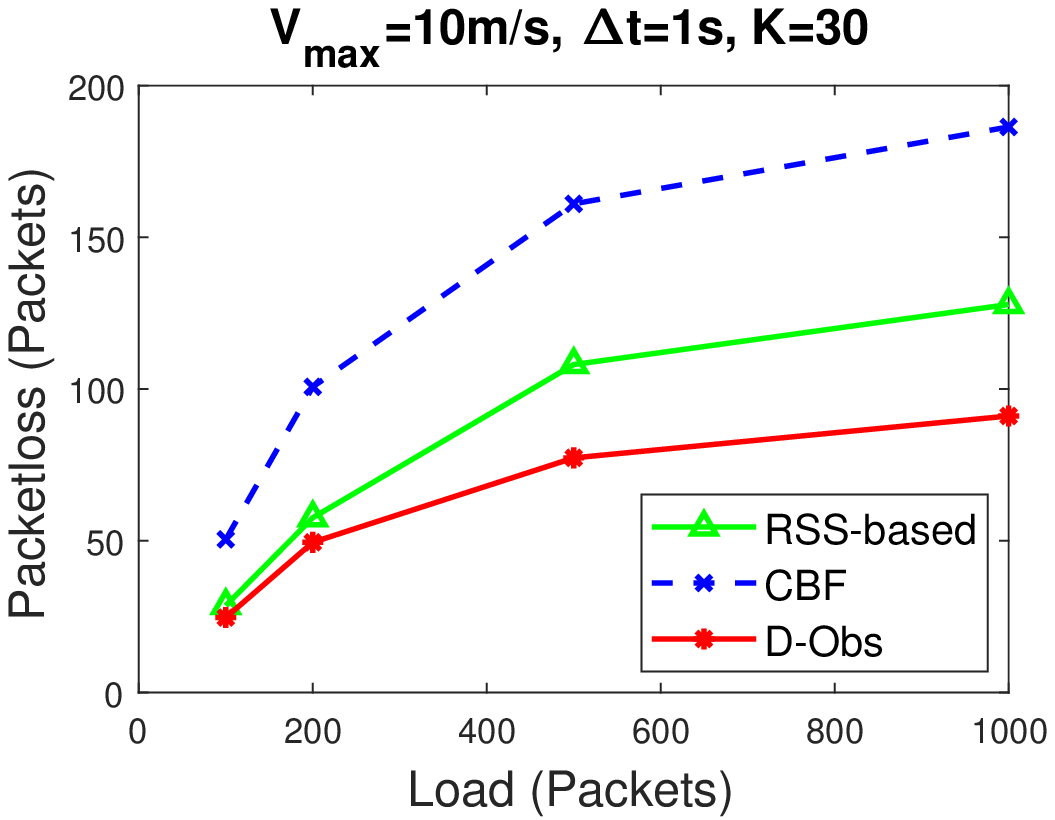}
			\caption{Load vs  avg. packetloss. }
			\label{fig:result5}
		\end{minipage}\hspace{0.5em}
		\begin{minipage}[b]{.24\textwidth}
			\includegraphics[width=1\textwidth]{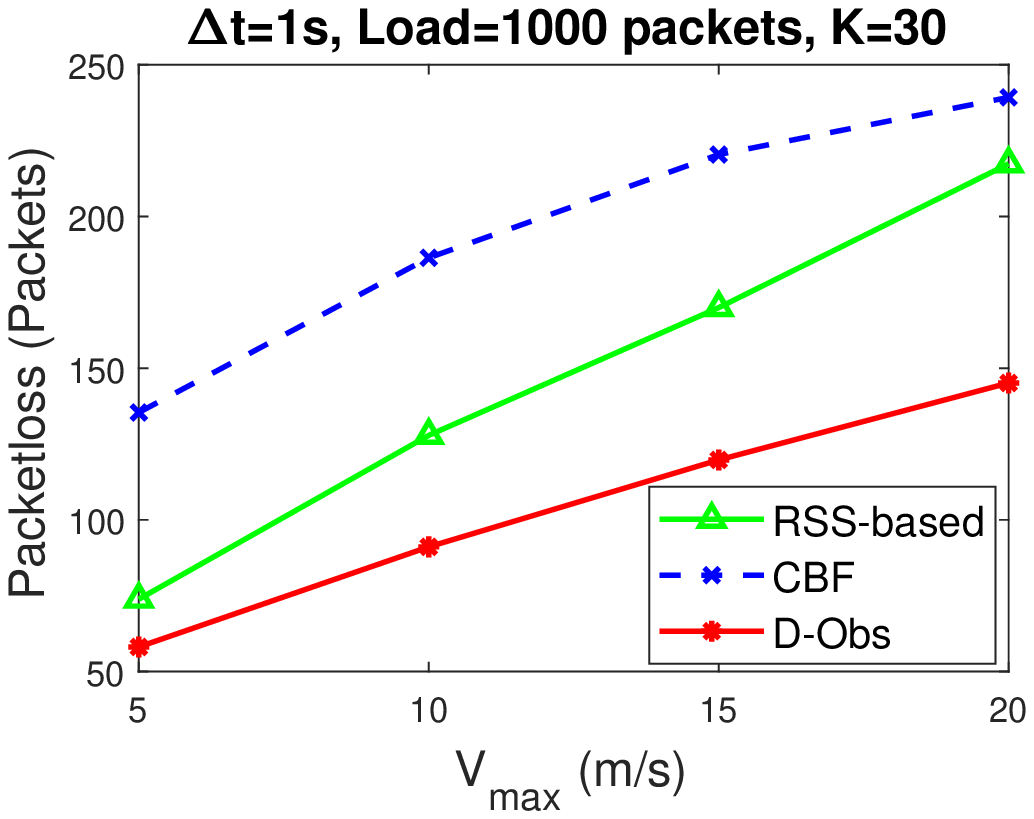}
			\caption{$V_{max}$ vs  avg. packetloss.  }
			\label{fig:result6}
		\end{minipage}\hspace{0.5em}
		\begin{minipage}[b]{.24\textwidth}
			\includegraphics[width=1\textwidth]{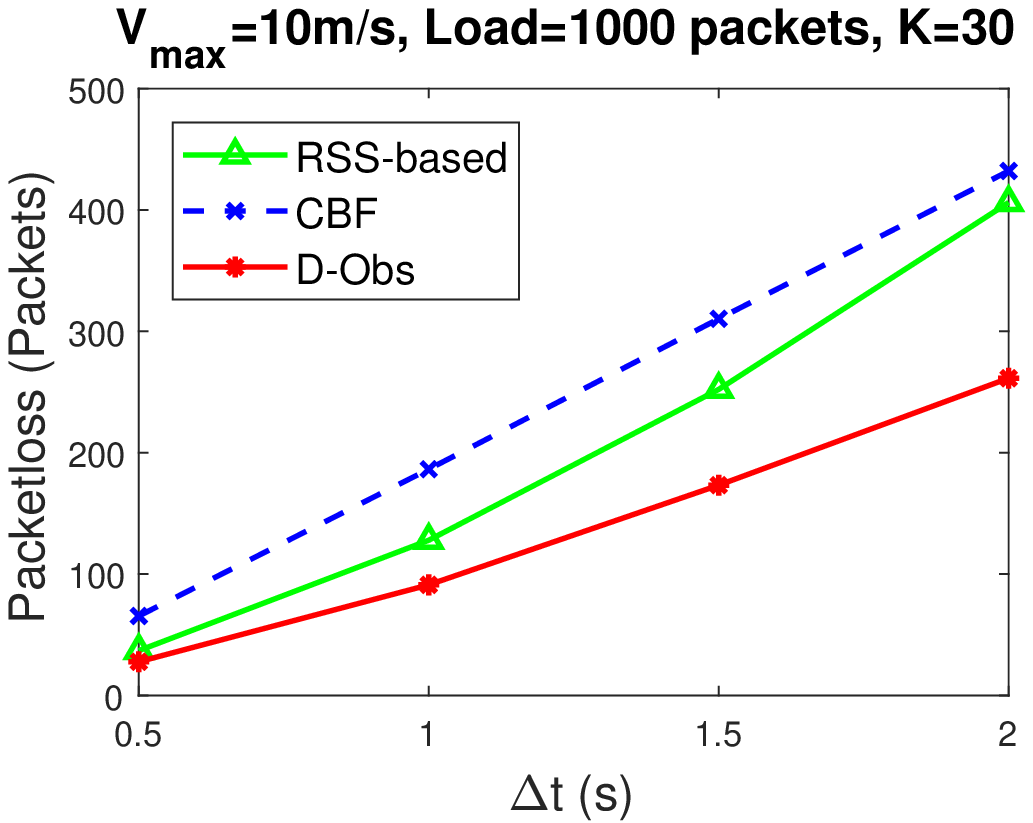}
			\caption{$\Delta t$ vs avg. packetloss. }
			\label{fig:result8}
		\end{minipage}
	\end{figure*}

	{\it Experimental Results \& Analysis}: 
	We have written our own C++ custom code and run them on a \texttt{GNU} $4.8$ compiler on Intel core $i7$ machine using the simulation environment mentioned in previous section and taken the average of the results of about $10000$ runs.
	
	Figure \ref{fig:result1} depicts the effect of varying $K$ on average throughput keeping other parameters fixed as mentioned in the figure. Average throughput decreases rapidly for \texttt{RSS} and \texttt{CBF} based approaches as compared to \texttt{D-Obs} algorithm, because both of them neither consider mobility nor obstacle and forward packets solely based on signal strength and distance from destination respectively. As a result the packet loss is also very high in both \texttt{RSS} and \texttt{CBF} based approaches as compared to \texttt{D-Obs} algorithm as shown in figure \ref{fig:result4}. Packet loss in \texttt{CBF} approach is higher than \texttt{RSS} because it chooses a relay based on its closer distance to the destination in which case the chosen relay \texttt{UE} can be far from the transmitting \texttt{UE} which may increase the chance of blockage from static \& dynamic obstacles. Also signal strength could be low causing a high end to end delay if packet loss occurs.
	
	Figure \ref{fig:result2} depicts the effect of varying network load on average throughput keeping other parameters fixed as mentioned in the figure. Here again average throughput for both \texttt{RSS} and \texttt{CBF} based approaches are lower as compared to \texttt{D-Obs} algorithm, because of the same reasons as mentioned above. As the number of packets to be sent increases, the chance of packets loss also increases due to mobility as well as obstacles which is shown in figure \ref{fig:result5}.
		
	Figure \ref{fig:result3} depicts the effect of varying $V_{max}$ on average throughput keeping other parameters fixed as mentioned in the figure. Here as the speed increases, the performance  of \texttt{RSS} and \texttt{CBF} deteriorate more rapidly as compared to \texttt{D-Obs} because with higher speed, nodes can move longer distance giving more chance for static and dynamic obstacle to interfere with them. Higher speed also causes more packet loss due to mobility as nodes may go out of range of each other quickly. The corresponding packet loss graph is shown in figure \ref{fig:result6}.
	
	Figure \ref{fig:result7} depicts the effect of varying $\Delta t$ on average throughput keeping other parameters fixed as mentioned in the figure. Here also, we can see that increasing  $\Delta t$ results in poor performance of  \texttt{RSS} and \texttt{CBF} compared to \texttt{D-Obs}. This is because, with higher $\Delta t$, nodes can move longer distance causing increase in chances of blockage by static and dynamic obstacles which in turn causes more packet loss. Also longer distance may cause packet loss due to mobility. The packet loss graph is shown in figure \ref{fig:result8}.
	\section{Conclusion}  \label{conclusion}
We formulated the problem of relay selection by capturing the effects of both obstacles and node's mobility. We optimized throughput by taking care of packet loss and average delay. To capture the motion of dynamic obstacles, we leveraged the radar employed with base station which would detect them with certain probability. Later we used geometrical analysis to derive unique solutions for computing the best relay node. In simulations we have shown the effects of both obstacles and node's mobility on throughput as well as packet loss. Results show that \texttt{D-Obs} outperforms other classical algorithms by appropriately capturing the effects of obstacles and node's mobility. 
	
	\bibliography{ref} 
	\bibliographystyle{ieeetr}

\end{document}